\documentclass[useAMS,usenatbib]{mn2e}

\newcommand{\feh}{\mbox{[Fe/H]}}
\newcommand{\teff}{\mbox{$T_{\rm eff}$}}
\newcommand{\logg}{\mbox{$\log g$}}
\newcommand{\vsini}{\mbox{$v \sin i$}}

\newcommand{\kms}{\mbox{km\,s$^{-1}$}}
\newcommand{\ms}{\mbox{m\,s$^{-1}$}}

\newcommand{\mplanet}{\mbox{$M_{\rm pl}$}}
\newcommand{\rplanet}{\mbox{$R_{\rm pl}$}}
\newcommand{\densplanet}{\mbox{$\rho_{\rm pl}$}}
\newcommand{\mjup}{\mbox{$M_{\rm Jup}$}}
\newcommand{\rjup}{\mbox{$R_{\rm Jup}$}}
\newcommand{\densjup}{\mbox{$\rho_{\rm Jup}$}}
\newcommand{\mstar}{\mbox{$M_{*}$}}
\newcommand{\rstar}{\mbox{$R_{*}$}}
\newcommand{\densstar}{\mbox{$\rho_*$}}
\newcommand{\msol}{\mbox{$M_\odot$}}
\newcommand{\rsol}{\mbox{$R_\odot$}}
\newcommand{\denssol}{\mbox{$\rho_\odot$}}
\newcommand{\secos}{\mbox{$\sqrt{e} \cos \omega$}} 
\newcommand{\sesin}{\mbox{$\sqrt{e} \sin \omega$}} 
\newcommand{\ecos}{\mbox{$e \cos \omega$}} 
\newcommand{\esin}{\mbox{$e \sin \omega$}} 
\newcommand{\svcos}{\mbox{$\sqrt{\vsini} \cos \lambda$}} 
\newcommand{\svsin}{\mbox{$\sqrt{\vsini} \sin \lambda$}} 
\newcommand{\fhband}{\mbox{$\Delta F_{1.6}$}} 
\newcommand{\fnbtwo}{\mbox{$\Delta F_{2.09}$}} 
\newcommand{\iracone}{\mbox{$\Delta F_{3.6}$}} 
\newcommand{\iractwo}{\mbox{$\Delta F_{4.5}$}} 
\newcommand{\iracthree}{\mbox{$\Delta F_{5.8}$}} 
\newcommand{\iracfour}{\mbox{$\Delta F_{8.0}$}}

\def\teqlf{$T_{\rm P,A=0,f}$}
\def\teqlfull{$T_{\rm P,A=0,f=1}$}
\def\teqlday{$T_{\rm P,A=0,f=2}$}
\def\teqlinst{$T_{\rm P,A=0,f=8/3}$}

\def\rhk{$\log R'_{\rm HK}$}

\newcommand{\aap}{A\&A}
\newcommand{\apj}{ApJ}
\newcommand{\aj}{AJ}
\newcommand{\apjl}{ApJ}
\newcommand{\mnras}{MNRAS}
\newcommand{\apjs}{ApJS}
\newcommand{\nat}{Nature}
\newcommand{\pasp}{PASP}

\usepackage{graphicx}
\usepackage{amssymb}
\usepackage{amsmath}
\usepackage{url}

\title[Thermal emission at 3.6--8 \micron\ from WASP-19b]
{Thermal emission at 3.6--8 \micron\ from WASP-19b: a hot Jupiter without 
a stratosphere orbiting an active star}

\author[D.~R.~Anderson et al.]
{D.~R.~Anderson,$^1$\thanks{dra@astro.keele.ac.uk} 
A.~M.~S.~Smith,$^1$ 
N.~Madhusudhan,$^2$ 
P.~J.~Wheatley,$^3$ 
\newauthor
A.~Collier~Cameron,$^4$ 
C.~Hellier,$^1$ 
C.~Campo,$^5$ 
M.~Gillon,$^6$ 
J.~Harrington,$^5$ 
\newauthor
P.~F.~L.~Maxted,$^1$ 
D.~Pollacco,$^7$
D.~Queloz,$^8$ 
B.~Smalley,$^1$ 
A.~H.~M.~J.~Triaud$^8$ 
\newauthor
and R.~G.~West$^{9}$\\
$^1$Astrophysics Group, Keele University, Staffordshire ST5 5BG, UK\\
$^2$Department of Astrophysical Sciences, Princeton University
    Princeton, NJ 08544, USA\\
$^3$Department of Physics, University of Warwick, Coventry, CV4 7AL, UK\\
$^4$School of Physics and Astronomy, University of St. Andrews, North Haugh, Fife, KY16 9SS, UK\\
$^5$Planetary Sciences Group, Department of Physics, University of Central Florida, Orlando, FL 32816-2385, USA\\
$^6$Institut d'Astrophysique et de G\'eophysique,  Universit\'e de Li\`ege,  All\'ee du 6 Ao\^ut, 17,  Bat.  B5C, Li\`ege 1, Belgium\\
$^7$Astrophysics Research Centre, School of Mathematics \& Physics, Queen's University, University Road, Belfast, BT7 1NN, UK\\
$^8$Observatoire de Gen\`eve, Universit\'e de Gen\`eve, 51 Chemin des Maillettes, 1290 Sauverny, Switzerland\\
$^9$Department of Physics and Astronomy, University of Leicester, Leicester, LE1 7RH, UK\\
}

\begin{document}

\date{Accepted 2013 January 22. Received 2013 January 21; in original form 2011 December 21}

\pagerange{\pageref{firstpage}--\pageref{lastpage}} \pubyear{2012}

\maketitle

\label{firstpage}

\begin{abstract}
We report detection of thermal emission from the exoplanet WASP-19b at 3.6, 
4.5, 5.8 and 8.0 \micron. 
We used the InfraRed Array Camera on the Spitzer Space Telescope to observe 
two occultations of WASP-19b by its host star. 
We combine our new detections with previous measurements of WASP-19b's emission 
at 1.6 and 2.09 \micron\ to construct a spectral energy distribution of the 
planet's dayside atmosphere. 
By comparing this with model-atmosphere spectra, we find that the dayside 
atmosphere of WASP-19b lacks a strong temperature inversion. 
As WASP-19 is an active star (\rhk\ = $-4.50 \pm 0.03$), this finding supports 
the hypothesis of \citet{2010ApJ...720.1569K} that inversions are suppressed in 
hot Jupiters orbiting active stars. 
The available data are unable to differentiate between a carbon-rich and an 
oxygen-rich atmosphere. 
\end{abstract}


\section{Introduction}
By observing the occultation of an exoplanet by its host star, we can measure 
the emergent flux from the planet's dayside atmosphere.
Such measurements are challenging due to the low planet-to-star flux ratio (for 
the best cases, typically a few tenths of one per cent in the near infrared) 
and sources of noise both instrumental and stellar in origin 
\citep[e.g.: ][]{2008ApJ...673..526K, 2011MNRAS.416.2096S}.
To date, most such measurements 
\citep[including the first,][]{2005ApJ...626..523C,2005Natur.434..740D} 
have been made with the {\it Spitzer Space Telescope}, though ground-based 
facilities are now an essential complement 
\citep[e.g.:][]{2009A&A...493L..35D,2009A&A...493L..31S}. 

With a single occultation we can measure the corresponding brightness 
temperature and determine the eccentricity of the planet's orbit 
\citep[e.g.][]{2005ApJ...626..523C}, which is necessary for the accurate 
determination of the stellar and planetary radii in transiting systems 
\citep{2012MNRAS.422.1988A} and is important for studies of the formation and 
tidal inflation of short-period, giant planets 
\citep[e.g.:][]{2011ApJ...742...72N,2010ApJ...713..751I}.

With photometric measurements at various wavelengths, we can construct a 
spectral energy distribution of the planet's dayside atmosphere. 
From this we can infer properties such as the planetary albedo, the dayside 
energy budget and the efficiency of dayside-to-nightside energy redistribution 
\citep[e.g.][]{2005ApJ...632.1132B}. 
As the atmospheric depth probed depends on the molecular opacity in the 
observation band, a spectrum can describe an atmosphere's vertical temperature 
structure. 
The observation that some planet atmospheres exhibit strong temperature 
inversions, or stratospheres, and others do not 
\citep[see][for a summary]{2010ApJ...720.1569K} 
led to a suggestion that inversions are present in atmospheres hot enough to 
maintain high-opacity absorbers in the gas phase in the
upper atmosphere \citep[e.g.][]{2008ApJ...678.1419F}. 
However, this was challenged by recent contrary results 
\citep{2008ApJ...684.1427M,2010ApJ...711..374F} and theory 
\citep[e.g.][]{2009ApJ...699.1487S}. 
As an alternative, \citet{2010ApJ...720.1569K} suggested that those planets 
orbiting chromospherically active stars lack inversions because the associated 
high UV flux destroys the high-opacity, high-altitude compounds that would 
otherwise induce inversions. 
This hypothesis is based on a small sample and further measurements 
across a wider parameter space are vital to test it. Specifically, there is 
a paucity of measurements for planets orbiting active stars. 
We suggest composition could be a key factor, with low-metallicity planets 
lacking high concentrations of the high-opacity absorbers, such as TiO and 
sulphur, thought to be responsible for inversions 
\citep[e.g.][]{2008ApJ...678.1419F,2009ApJ...701L..20Z}. 

As the spectral coverage increases, so too does the information about an 
atmosphere that we can discern. 
\citet{2011Natur.469...64M} used seven measurements of the 
emission of WASP-12b \citep{2011AJ....141...30C,2011ApJ...727..125C}  
to show that the dayside atmosphere is the first known to be carbon-dominated. 
Thus we are entering an era in which we can make statistically-sound 
inferences about the composition of exoplanet atmospheres. 
Further, \citet{2011Natur.469...64M} demonstrated that the planet lacks a 
prominent thermal inversion and has very inefficient day-night energy 
circulation. 

In this paper, we present {\it Spitzer} measurements of WASP-19b's dayside 
thermal emission at 3.6, 4.5, 5.8 and 8.0 \micron. 
Discovered by the Wide Angle Search for Planets 
\citep{2006PASP..118.1407P,2010ApJ...708..224H}, the shortest-period hot 
Jupiter, WASP-19b, is a 1.17-\mjup\ planet in a near-circular, 19-hr orbit 
around a G8V star. 
\citet{2011ApJ...730L..31H} showed the planet's orbital axis to be aligned with 
the spin axis of its host star. 
The planet's thermal emission was previously measured at 1.6 and 2.09 \micron\ 
\citep{2010A&A...513L...3A, 2010MNRAS.404L.114G}, thus we bring the measurement 
tally to six bands. We use all six thermal emission measurements to 
characterise the planet's atmosphere and refine the system parameters by 
combining these data with pre-existing photometry of the transit and 
radial-velocity data. 

\section{New observations}

We observed two occultations of the planet WASP-19b by its host star WASP-19 
(2MASS 15595095$-$2803422, $K_{s}=10.22$) with {\it Spitzer} 
\citep{2004ApJS..154....1W} during UT 2009 January 29 and UT 2009 March 22.
On each date, we employed the Infrared Array Camera 
\citep[IRAC, ][]{2004ApJS..154...10F} in full array mode 
($256\times256$\,pixels, 1.2~\arcsec\ pix$^{-1}$).
During the first occultation, we measured the WASP-19 system simultaneously in 
the 4.5- and 8.0-\micron\ channels (respectively, channels 2 and 4) for a 
duration of 3.2 h. 
We measured the second occultation simultaneously in the 3.6 and 
5.8~\micron\ channels (respectively, channels 1 and 3) for a duration of 3.0 h. 

Prior to the first occultation, we used the emission nebula NGC 7538 in Cepheus 
to `pre-flash' the target position on the detector arrays for 0.5 hr. 
This was intended to reduce or remove the known illumination-history 
dependence of the gain response of the 8.0-\micron\ detector 
\citep[e.g.][and references therein]{2008ApJ...673..526K}.
The 5.8-\micron\ detector is known to suffer a similar issue, but pre-flashing 
this array was not permitted due to the detrimental effect it is known to have 
on the array. 
Instead, we attempted to stabilize the array by 
observing the target for an extra hour prior to the occultation. 
As this `pre-stare' was performed as a separate observation request to the 
occultation observation, the target was reacquired between the two. 
As such, the pre-stare observation must be treated as a separate dataset with 
its own systematics and so are of little use in determining the occultation 
depth. 

Using an effective integration time of 10.4~s, we obtained 876 and 
840~images, respectively, for the first and second occultations. 
In each dataset, we see a small, periodic, pointing wobble ($P\approx1$ hr), 
thought to be caused by the thermal cycling of an on-board battery 
heater\footnote{http://ssc.spitzer.caltech.edu/warmmission/news/21oct2010memo.pdf}. 
There is also a very small drift of the target position over the span of each 
dataset.

We used the images calibrated by the standard {\it Spitzer} pipeline (version 
S18.7.0) and delivered to the community as Basic Calibrated Data (BCD). 
Our method is essentially the same as we presented in 
\citet{2011MNRAS.416.2108A}, to which we refer the reader for further 
information. 
For each image we converted flux from MJy~sr$^{-1}$ to electrons and then 
used {\sc iraf} to perform aperture photometry for WASP-19, using circular 
apertures with a range of radii: 1.5--6~pixels for the 3.6 \micron\ and 4.5 
\micron\ data and 1--5~pixels for the 5.8 \micron\ and 8.0 \micron\ data. 
The apertures were centred by fitting a Gaussian profile on the target. 
The sky background was measured in an annulus extending from 8 to 12 pixels 
from the aperture centre, and was subtracted from the flux measured within the 
on-source apertures. 
We estimated the photometric uncertainty as the quadrature addition of the 
uncertainty in the sky background (estimated as the standard deviation of the 
flux in the sky annulus) in the on-source aperture, 
the read-out noise, and the Poisson noise of the total background-subtracted 
counts within the on-source aperture.
We calculated the mid-exposure times in the HJD (UTC) time system from the 
MHJD\_OBS header values, which are the start times of the DCEs (Data Collective 
Events), by adding half of a DCE duration (FRAMTIME).

The choice of aperture radius for each dataset was a compromise between 
maximising the signal-to-noise of the measurements and, from fits to all 
available data (see Section 3), minimising the residual scatter of the 
lightcurve. 
Each consideration suggested very similar optimal apeture radii and we 
adopted 2.7 pix for the 3.6- and 4.5-\micron\ data and 2.5 pix for the 5.8- 
and 8.0-\micron\ data. 
For each dataset we found that the variation in the fitted occultation depth 
(see Section 3) was much smaller than 1\,$\sigma$ over a wide range of aperture 
radii. 

\begin{figure*}
\begin{center}
$\begin{array}{ccc}
\includegraphics[scale=0.98]{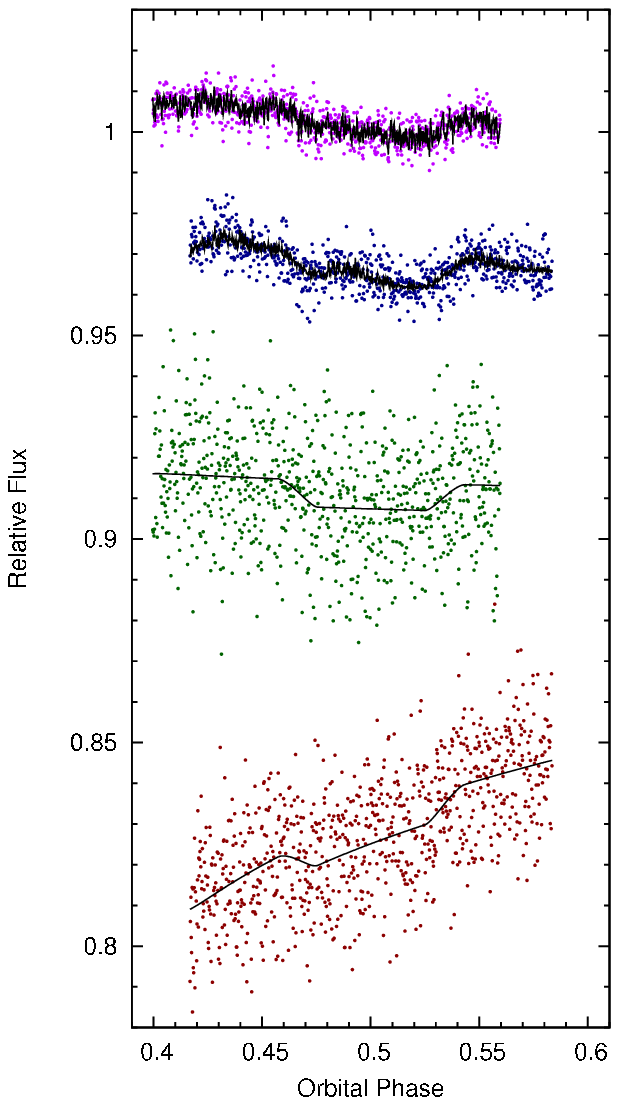} &
\includegraphics[scale=0.98]{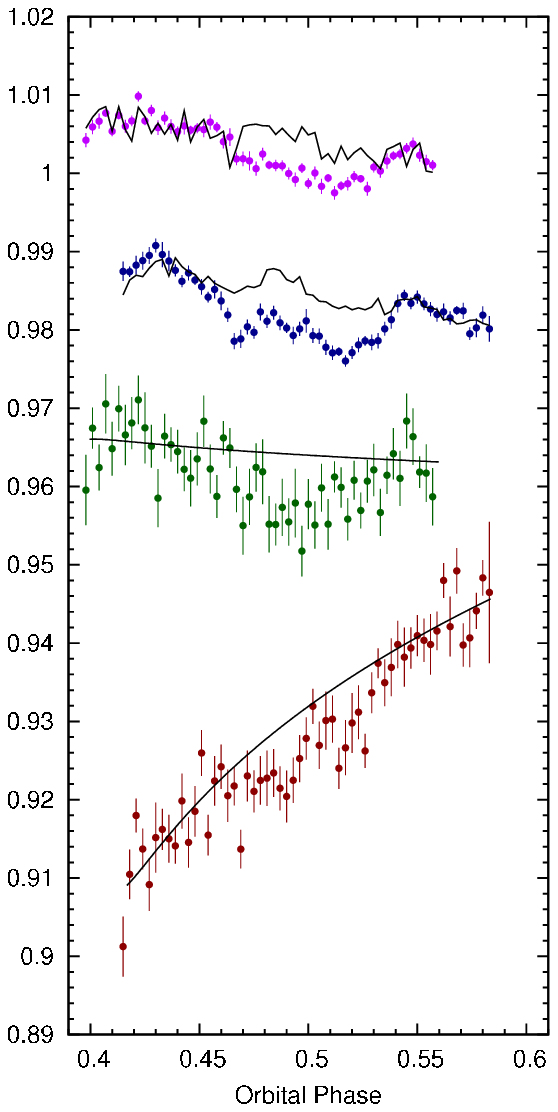} &
\includegraphics[scale=0.98]{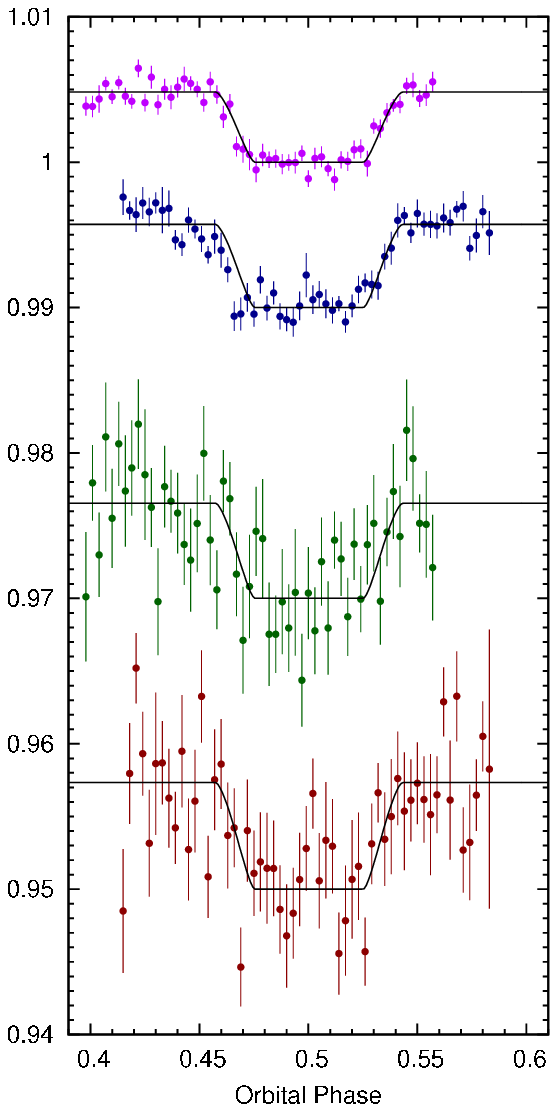}
\end{array}$
\caption{
In each of the above three plots, from top to bottom, the data were taken at 
3.6, 4.5, 5.8 and 8.0 \micron. 
Relative flux offsets were applied to datasets for clarity. 
{\bf\em Left}: Raw {\it Spitzer} data with the best-fitting trend and 
occultation models superimposed.
{\bf\em Middle}: The same data binned in phase ($\Delta \phi=0.003 \sim 3.4$ 
min) with the best-fitting trend models superimposed.
{\bf\em Right}: The binned data after dividing by the best-fitting trend models, 
and with the best-fitting occultation models superimposed.
The error bar on each binned measurement in the panels in the middle and on the 
right is the standard deviation of the points within the bin.
\label{fig:spitz}}
\end{center}
\end{figure*}

Some groups choose to reject a portion of data at the beginning of each 
observation, citing as justification either the settling of the spacecraft (not 
seen in our data) or an improvement in the fit; we found no reason to do this. 
We rejected any flux measurement that was discrepant with the median of its 20 
neighbors (a window width of 4.4 min) by more than four times its theoretical 
error bar.
We also performed a rejection on target position. For each image and for the 
{\it x} and {\it y} detector coordinates separately, we computed the difference 
between the fitted target 
position and the median of its 20 neighbors. For each dataset, we then 
calculated the standard deviation, $\sigma$, of these {\it median differences} 
and rejected any points discrepant by more than 4\,$\sigma$.
The numbers of points rejected on flux and target position for each dataset are 
displayed in Table~\ref{tab:rej}.
According to the IRAC handbook, each IRAC array receives approximately 1.5 
solar-proton and cosmic-ray hits per second, with $\sim$2 pixels per hit 
affected in channels 1 and 2, and $\sim$6 pixels per hit affected in channels 3 
and 4, and the cosmic ray 
flux varies randomly by up to a factor of a few over time scales of minutes. 
Thus, the average probability per exposure that pixels within the stellar 
aperture will be affected by a cosmic ray hit is 1.3 per cent for channels 1 and 
2 and 3.2 per cent for channels 3 and 4, which is in good agreement with the 
portion of frames that we rejected. 
These probabilities are likely to be underestimates as we calculated them using 
partial pixels and neglecting the effect of hits within the sky annuli. 
For an unknown reason, a greater portion of channel 2 images were rejected due 
to jumps of the target position, particularly in the {\it x} direction. 
The post-rejection data are displayed raw and binned in the first and second  
panels respectively of Figure~\ref{fig:spitz}. 

\begin{table}
\begin{center}
\caption{Number of points rejected per dataset per criterion\label{tab:rej}}
\begin{tabular}{lrrrr}
\hline
Dataset			& Flux	& {\it x}-pos	& {\it y}-pos	& Total (\%)\\
\hline
2009 Mar 22 / 3.6 \micron	&  8	&  1	&  3		&  11 (1.3)\\
2009 Jan 29 / 4.5 \micron	& 11	& 41	& 11		&  51 (5.8)\\
2009 Mar 22 / 5.8 \micron	& 11	&  9	&  7		&  21 (2.5)\\
2009 Jan 29 / 8.0 \micron	&  4	&  8	&  9		&  17 (1.9)\\
\hline
\end{tabular}
\end{center}
\end{table}

\section{Data analysis}

\subsection{Data and model}

We performed a global determination of the system parameters incorporating: our 
new {\it Spitzer} occultation photometry; 
the HAWK-I $H$-band and 2.09-\micron\ 
occultation lightcurves obtained, respectively, 
by \citet{2010A&A...513L...3A} and \citet{2010MNRAS.404L.114G}; 
the 34 CORALIE radial-velocity (RV) measurements listed in 
\citet{2010ApJ...708..224H}; 
the 36 HARPS RVs, obtained through a transit, and the 3 CORALIE RVs given 
in \citet{2011ApJ...730L..31H}; 
the LCOGT FTS $z$-band transit lightcurve from \citet{2010ApJ...708..224H}; 
and the ESO NTT $r$-band transit lightcurve presented in 
\citet{2011ApJ...730L..31H}. 
We did not include the three seasons of WASP survey photometry presented in 
\citet{2010ApJ...708..224H}. Rather, we placed a Bayesian Gaussian prior on the 
epoch of mid-transit, $T_{\rm c}$, using the epoch given in 
\citet{2011ApJ...730L..31H}: $T_{\rm c} = 2455168.96801 \pm 0.00009$ HJD.
Thus our analyses completed quicker and the shape of the transit was 
determined using only the high-S/N photometry. This can be preferable as 
photometry from surveys such as WASP is prone to dilution and, depending on 
which detrending algorithm is used, the transit depth can be suppressed. 
We decorrelated each transit lightcurve with a linear function of time. 
The HAWK-I $H$-band data were partitioned and detrended as in 
\citet{2010A&A...513L...3A} and the HAWK-I 2.09-\micron\ lightcurve was 
decorrlelated with a linear function of time as in \citet{2010MNRAS.404L.114G}. 
These data were used as input into an adaptive Markov-chain Monte Carlo (MCMC) 
algorithm \citep{2007MNRAS.380.1230C,2008MNRAS.385.1576P, 2010A&A...516A..33E}; 
see \citet{2011A&A...534A..16A} for a description of the current version of 
our code. 
Such an analysis, incorporating all available data, is necessary to take account 
of the cross-dependancy of system parameters and to make a reliable assessment 
of their uncertainties. 
We partitioned the RV data by spectrograph so as to allow for an instrumental 
offset and for a potential specific stellar activity level during the 
short-baseline HARPS observations. 

The MCMC proposal parameters we used are: $T_{\rm c}$, $P$, 
$(\rplanet/\rstar)^2$, $T_{14}$, $b$, $K_{\rm 1}$, \teff, \feh, \secos, \sesin, 
\svcos, \svsin, \fhband, \fnbtwo, \iracone, \iractwo, \iracthree, \iracfour\ 
and $t_{\rm off}$.
See Section~\ref{sec:spdata} for a definition of $t_{\rm off}$ and 
Table~\ref{tab:sys-params} for definitions of the other parameters. 
At each step in the MCMC procedure, each proposal parameter is perturbed from
its previous value by a small, random amount. 
From the proposal parameters, model light and RV curves are generated and 
$\chi^{2}$ is calculated from their comparison with the data. 
A step is accepted if $\chi^{2}$ (our merit function) is lower than for the 
previous step, and a step with higher $\chi^{2}$ is accepted with probability 
$\exp(-\Delta \chi^{2}/2)$. 
In this way, the parameter space around the optimum solution is thoroughly 
explored.
The value and uncertainty for each parameter are respectively taken as the 
median and central 68.3 per cent confidence interval of the parameter's 
marginalised posterior probability distribution 
\citep[e.g.][]{2006ApJ...642..505F}.

\subsection{Spitzer data}
\label{sec:spdata}

IRAC uses an InSb detector to observe at 3.6 and 4.5 \micron, and the 
measured flux exhibits a strong correlation with the position of the target 
star on the array. 
This effect is due to the inhomogeneous intra-pixel sensitivity of the detector 
and is well-documented 
\citep[e.g.][and references therein]{2008ApJ...673..526K}.
Following \citet{2008ApJ...686.1341C} we modelled this effect as a 
quadratic function of the sub-pixel position of the PSF centre, 
with the addition of a cross-term to permit rotation \citep{2009ApJ...699..478D}
and a linear term in time:

\begin{equation}
\label{eqn:ch2}
df = a_0 + a_xdx + a_ydy + a_{xy}dxdy + a_{xx}dx^2 + a_{yy}dy^2 + a_tdt
\end{equation}

\noindent where $df = f - \hat{f}$ is the stellar flux relative to its weighted 
mean, $dx = x - \hat{x}$ and $dy = y - \hat{y}$ are the coordinates of the PSF 
centre relative to their weighted means, $dt$ is the time elapsed since the 
first observation, and $a_0$, $a_x$, $a_y$, $a_{xy}$, $a_{xx}$, 
$a_{yy}$ and $a_t$ are coefficients.

\begin{table}
\caption{Trend model parameters and coefficients} 
\label{tab:coeffs} 
\begin{tabular*}{0.5\textwidth}{@{\extracolsep{\fill}}lcccc} 
\hline 
 		& 3.6 \micron			& 4.5 \micron			& 5.8 \micron			& 8.0 \micron \\
\hline
\\
$\hat{f}$	& 151194.55			& 80580.23			& 12693.57			& 18991.68			\\
$\hat{x}$	& 31.93				& 24.50				& 25.99				& 25.44				\\
$\hat{y}$	& 24.91				& 25.98				& 25.16				& 23.63				\\
$a_0$		& $-24 \pm 16$			& $89.7 \pm 5.0$		& $-95.2^{+26}_{-22}$		& $1569^{+142}_{-116}$		\\
$a_x$		& $6566 \pm 75$			& $-433 \pm 36$			& ---				& ---				\\
$a_y$		& $9278 \pm 49$			& $4062 \pm 32$			& ---				& ---				\\
$a_{xy}$	& $6040 \pm 3280$		& $-7860 \pm 4150$		& ---				& ---				\\
$a_{xx}$	& $-6330 \pm 1230$		& $23000 \pm 880$		& ---				& ---				\\
$a_{yy}$	& $-22500 \pm 1980$		& $30360 \pm 3810$		& ---				& ---				\\
$a_t$		& $1050 \pm 270$		& $-2203 \pm 57$		& ---				& ---				\\
$a_1$		& ---				& ---				& $-52^{+21}_{-11}$		& $800^{+110}_{-86}$		\\
$a_2$		& ---				& ---				& $-5.9^{+4.1}_{-1.4}$		& $78^{+18}_{-13}$		\\
$t_{\rm off}$/min	& ---				& ---			& $10.6^{+11.4}_{-8.1}$		& $15.3^{+10.1}_{-7.4}$		\\
\\
\hline
\end{tabular*}
\end{table}

IRAC uses a SiAs detector to observe at 5.8 and 8.0 \micron, and its response is 
thought to be homogeneous, though another systematic affects the photometry. 
This effect is known as the `ramp' because it causes the gain to 
increase asymptotically over time for every pixel, with an amplitude depending 
on a pixel's illumination history 
\citep[e.g.][and references therein]{2008ApJ...673..526K}. 
Again following \citet{2008ApJ...686.1341C}, we modelled this ramp as a 
quadratic function of $\ln(dt)$:

\begin{equation}
\label{eqn:ch4}
df = a_0 + a_1\ln(dt+t_{\rm off}) + a_2(\ln(dt+t_{\rm off}))^2
\end{equation}

\noindent where $t_{\rm off}$ is a proposal parameter restricted to positive 
values. To prevent $t_{\rm off}$ 
from drifting to values greater than an hour or so, we place 
on it a Gaussian prior by adding a Bayesian penalty to our merit function 
($\chi^2$):

\begin{equation}
BP_{t_{\rm off}} = t_{\rm off}^2 / \sigma_{t_{\rm off}}^2
\end{equation}

\noindent where $\sigma_{t_{\rm off}}$ = 15 min.

A steep ramp is evident in the 8.0-\micron\ lightcurve (Figure~\ref{fig:spitz}, 
middle panel). 
Due to the large distance on the sky between the target and the pre-flash 
source, there was an 11-minute gap between the end of the pre-flash 
observations and the start of the target observations. 
It may be that the detector traps de-populated during this time, giving rise to 
the observed ramp that is more typical of observations without pre-flash.

In addition to Equation~\ref{eqn:ch4}, we tried trend functions with a variety 
of time dependencies: no time dependency; a linear-logarithmic time dependency 
(equivalent to setting $a_2=0$ in Equation~\ref{eqn:ch4}); 
a linear time dependency; and a quadratic time dependency. 
Each of these functions result in depths consistent within 1-$\sigma$ with the 
depths obtained using Equation~\ref{eqn:ch4}. 
For this reason and because Equation~\ref{eqn:ch4} has been shown to 
accurately describe the ramp in higher cadence, longer baseline datasets 
obtained with the SiAs detectors \citep[e.g.][]{2009ApJ...703..769K}, we adopt 
Equation~\ref{eqn:ch4} as our trend model for channels 3 and 4. 

We used singular value decomposition \citep{press_numerical_1992} to 
determine the trend model coefficients by linear least-squares minimization 
at each MCMC step, subsequent to division of the data by the eclipse model. 
The best-fitting trend models are superimposed on the binned photometry in the 
middle panel of Figure~\ref{fig:spitz}. 
Table~\ref{tab:coeffs} gives the best-fitting values for the trend model 
parameters and coefficients (Equations~\ref{eqn:ch2} and \ref{eqn:ch4}), 
together with their 1-$\sigma$ uncertainties.

\subsection{Photometric and RV noise}
\label{sec:noise}
We scaled the formal photometric error bars so as to obtain a reduced 
$\chi^{2}$ of unity, applying one scale factor per dataset. The aim was to 
properly weight each dataset in the simultaneous MCMC analysis and to obtain 
realistic uncertainties. 
The error bars of the FTS and the NTT photometry were multiplied, respectively, 
by 0.71 and 1.15. 
The scale factors for the error bars of the occultation photometry from IRAC 
channels 1, 2, 3 and 4 were, respectively, 1.02, 0.99, 1.16 and 1.04. 
Importantly, the error bars of the occultation photometry were not scaled when 
deciding which trend models or aperture radii to use. 
In \citet{2010A&A...513L...3A}, the HAWK-I $H$-band occultation data were split 
into eleven lightcurves according to offset and telescope repointing. 
Within a global analysis, each lightcurve was detrended individually and the 
error bars of each lightcurve was rescaled by its own factor. 
As the number of datapoints in each lightcurve is small (8--12 in the 6 
lightcurves obtained prior to repointing and 40--44 in the 5 lightcurves 
obtained post repointing), we here opted to rescale the error bars of all 
eleven lightcurves by the same factor, which was was 1.14, though we did still 
detrend each lightcurve separately. 
The error bars of the 2.09-\micron\ lightcurve presented in 
\citet{2010MNRAS.404L.114G} were scaled by 6.15. 
\citet{2010MNRAS.404L.114G} also found their uncertainties required a large 
scaling factor (6.17). They attributed this to variations in the 
inter-pixel sensitivity that likely resulted from their random dithering 
(radius = 30\arcsec) of the pointing between observations. 
Though our $H$-band observations were also made using HAWK-I 
\citep{2010A&A...513L...3A}, the corresponding scale factor was much closer to 
unity. This is because we dithered over a fixed pattern of six offsets and did 
not employ random jitter. 
We could thus produce one lightcurve per offset position and model each of 
their systematics independently. 
Current best practice is to avoid offsetting at all. 

\begin{figure}
\includegraphics[width=0.45\textwidth]{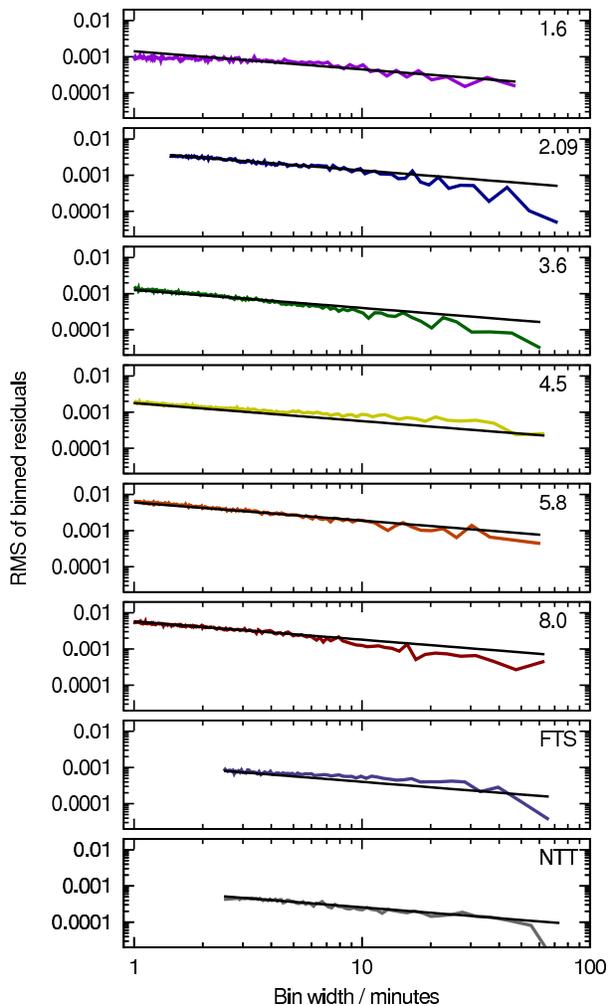}
\caption{
RMS of the binned residuals for, from top to bottom, the pre-existing HAWK-I 
occultation photometry at 1.6 \micron\ and 2.1 \micron, 
the new Spitzer occultation photometry at 3.6, 4.5, 5.8 and 8.0 \micron, 
and the pre-existing FTS and NTT transit photometry. 
The solid black lines, which are the rms of the unbinned 
data scaled by the square root of the number of points in each bin, show the 
white-noise expectation. 
The ranges of bin widths are appropriate for the datasets' cadences and 
durations. 
\label{fig:resids}}
\end{figure}

We assessed the presence of correlated noise in the occultation and transit data 
by plotting the rms of their binned residuals (Figure~\ref{fig:resids}). 
The 4.5-\micron\ occultation lightcurve and the FTS transit lightcurve each 
display a small amount of correlated noise on timescales of 5 min and longer. 

To obtain a reduced spectroscopic-$\chi^2$ of unity and to balance the different 
datasets in the MCMC, we added in quadrature a jitter of 14.1 \ms\ 
to the uncertainties of the CORALIE RVs and 6.9 \ms\ to the uncertainties of the 
HARPS RVs, as was done in \citet{2011ApJ...730L..31H}. 

\subsection{Time systems and light travel time}
The {\it Spitzer}, HAWK-I, FTS and NTT photometry are in the HJD (UTC) time 
system. The CORALIE and HARPS RVs are in the BJD (UTC) time system. 
The difference between BJD and HJD is less than 4~s and so is negligible for our 
purposes, and the timing information mostly comes from the photometry. 
Leap second adjustments are made to the UTC system to keep it close to 
mean solar time, so one should really use Terrestrial Time.
However, our observations span a short baseline (2008--2010), during which there 
was only one leap second adjustment, so our choice to use the UTC system has no 
impact. 

The occultation of WASP-19b occurs farther away from us than its transit does, 
so we made a first order correction for the light travel time. We calculated the 
light travel time between the beginning of occultation ingress and the beginning 
of transit ingress to be 15 s. We subtracted this from the mid-exposure times 
of the {\it Spitzer} and HAWK-I occultation photometry. 
For comparison, we measure the time of mid-occultation to a precision of 35 s.

\subsection{Results}
\begin{figure*}
\begin{centering}
\includegraphics[scale=0.95]{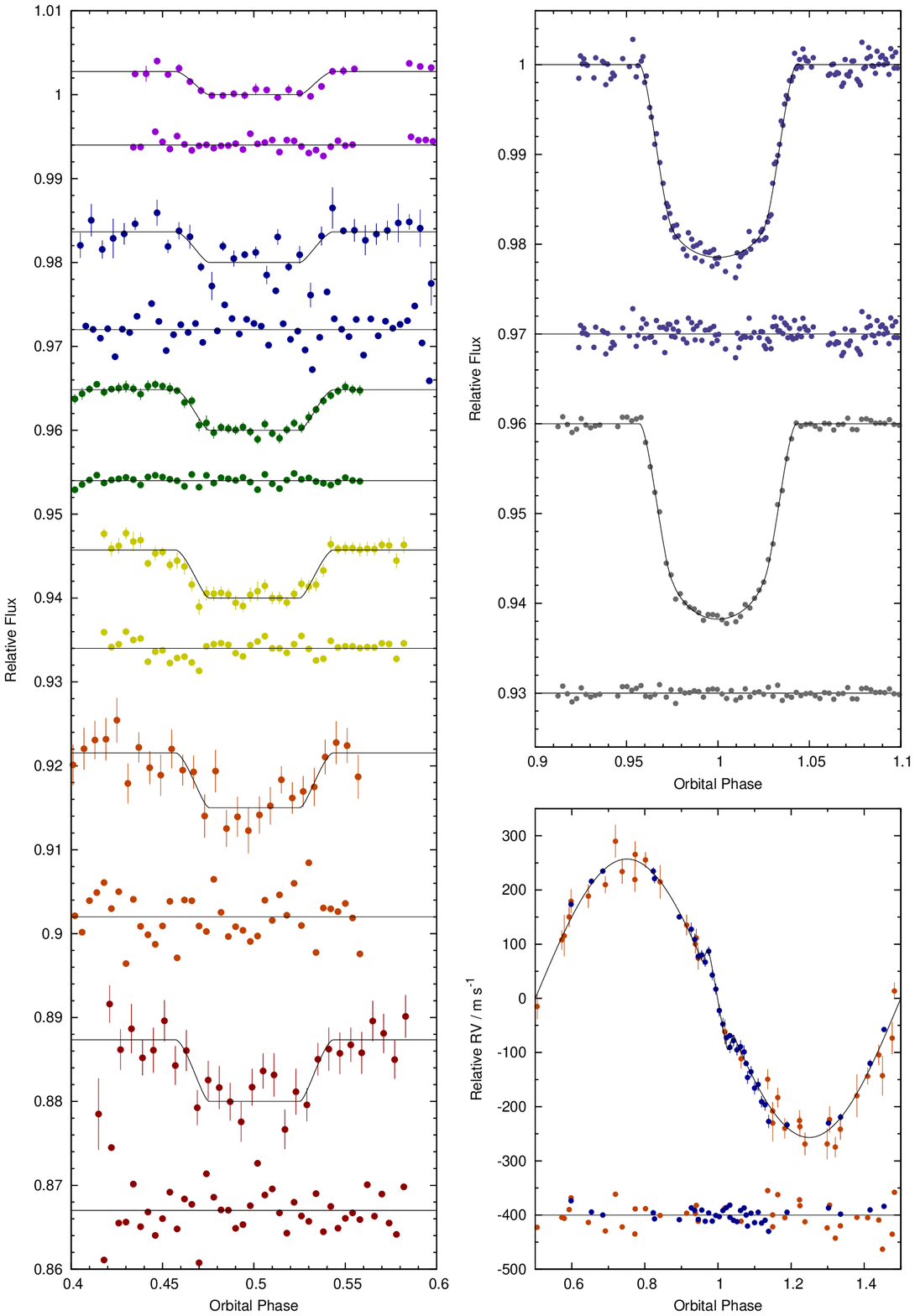}
\caption{
The results of our global analysis, which combines our new {\it Spitzer} 
occultation photometry with pre-existing transit photometry, ground-based 
occultation photometry and radial velocities. 
The models generated from the best-fitting parameter values of 
Table~\ref{tab:sys-params} are overplotted and the residuals about the models 
are plotted below each dataset. 
{\bf\em Left}: From top to bottom, HAWK-I occultations at 1.6 \micron\ 
\citep{2010A&A...513L...3A} and 2.09 \micron\ \citep{2010MNRAS.404L.114G}. 
The error bar on each binned measurement is the standard deviation of the 
points within the bin.
{\bf\em Top-right}: $z$-band transit lightcurve taken with FTS 
\citep[top;][]{2010ApJ...708..224H} and $r$-band transit lightcurve taken 
with NTT \citep[bottom;][]{2011ApJ...730L..31H}. 
Arbitrary offsets have been applied to the photometry plots for display 
purposes and 
{\bf\em Bottom-right}: Spectroscopic orbit and transit illustrated by CORALIE 
and HARPS data \citep{2010ApJ...708..224H,2011ApJ...730L..31H}. 
The measured systemic velocities (Table~\ref{tab:sys-params}) of each dataset 
have been subtracted.
\label{fig:mcmc}}
\end{centering}
\end{figure*}

Table~\ref{tab:sys-params} shows the median values and the 1-$\sigma$ 
uncertainties of the fitted proposal parameters and derived parameters from 
our final MCMC analysis. 
Figure~\ref{fig:spitz} shows the best-fitting trend and occultation models 
together with the raw and detrended {\it Spitzer} data. 
Table~\ref{tab:coeffs} gives the best-fitting values for the parameters of the 
trend models, together with their 1-$\sigma$ uncertainties.  
Figure~\ref{fig:mcmc} displays all the photometry and RVs used in the MCMC 
analysis, with the best-fitting eclipse and radial-velocity models superimposed.

\citet{2010A&A...513L...3A} measured an occultation depth of $0.259 \pm 0.045$ 
per cent from their $H$-band lightcurve. In analysing the same data, we obtained 
a similar depth of $0.276 \pm 0.044$ per cent. 
The minor difference is probably due to some combination of the slightly 
different manner in how the lightcurves' error bars were rescaled (see 
Section~\ref{sec:noise}) 
and the slight difference in the occultation ephemeris (the additional 
radial-velocity and occultation data result in a mid-occultation time 
at the time of the $H$-band observations $\sim$5 min earlier than found by 
\citet{2010A&A...513L...3A}).
The occultation depth of $0.366 \pm 0.067$ per cent that we derived from 
the 2.09-\micron\ lightcurve of \citet{2010MNRAS.404L.114G} is near-identical to the depth that they obtained ($0.366 \pm 0.072$ per cent).

We used the residual permutation or `prayer bead' method (e.g. \citealt{2007A&A...471L..51G}) as described in \citet{2012A&A...545A..93S} to assess the impact of any correlated noise present in the occultation lightcurves on our fitted occultation depths. The occultation depths derived from that analysis are consistent with the depths we measured from the non-permuted lightcurves and it is the latter that we adopt (Table~\ref{tab:permut}). 

We calculated the brightness temperatures that correspond to the measured 
occultation depths and present these in Table~\ref{tab:sys-params}. 
To calculate these, we defined the product of the planet-to-star area ratio and 
the ratio of the band-integrated planet-to-star flux densities, 
corrected for the wavelength-dependency of the transmission\footnote{For the 
HAWK-I measurements, the transmission of the atmosphere, telescope, instrument, 
and detector were accounted for by using the transmisson curve obtained from 
\url{http://www.eso.org/observing/etc/}. 
For the {\it Spitzer} measurements, the telescope throughput and detector 
quantum efficiency were accounted for by using the the full array average 
spectral response curves available at 
\url{http://irsa.ipac.caltech.edu/data/SPITZER/docs/}\\
\url{irac/calibrationfiles/spectralresponse/}.}, 
to be equal to the measured occultation depth (e.g., Charbonneau et al. 2005). 
We used a model spectrum of a star with the \teff, \logg\ and \feh\ values of 
Table~\ref{tab:sys-params} \citep{1999ApJ...512..377H}, normalised to 
reproduce the integrated flux of a black body with \teff~$ = 5475$\,K. 
The uncertainties in the brightness temperatures only take into account the 
uncertainties in the measured occultation depths.

To obtain reliable determinations of the occultation depths and orbital 
eccentricity, it is important that the time of mid-transit is known 
with accuracy and precision at the epochs the occultation data are obtained. 
The WASP photometry, which we used to place a prior on the transit ephemeris, 
span a two-year baseline of 2006 May to 2008 May and the FTS transit 
lightcurve was obtained in 2008 December. 
All occultation data were obtained soon after: during 2009 January to April. 
The NTT transit lightcurve, obtained in 2010 February, ensured a reliable 
transit ephemeris at the occultation epochs. 
We note, though, that the difference between the transit ephemeris presented 
herein and that presented in the discovery paper 
\citep[][i.e. without the NTT lightcurve]{2010ApJ...708..224H}, propagated to 
the occultation epochs, is a mere $\sim$20~s. 
The accuracy of the discovery-paper ephemeris is due to the long 
baseline of the WASP photometry and the high quality of the FTS lightcurve, 
which was obtained only months before the occultation data. 

\begin{table*} 
\caption{System parameters of WASP-19} 
\label{tab:sys-params} 
\begin{tabular}{llcl}
\hline 
Parameter & Symbol & Value & Unit\\ 
\hline 
\\
Orbital period & $P$ & $0.78883951 \pm 0.00000032$ & d\\
Epoch of mid-transit (HJD, UTC) & $T_{\rm c}$ & $2455183.16711 \pm 0.000068$ & d\\
Transit duration (from first to fourth contact) & $T_{\rm 14}$ & $0.06549 \pm 0.00035$ & d\\
Duration of transit ingress $\approx$ duration of transit egress & $T_{\rm 12} \approx T_{\rm 34}$ & $0.01346 \pm 0.00052$ & d\\
Planet-to-star area ratio & (\rplanet/\rstar)$^2$ & $0.02050 \pm 0.00024$ & \\
Impact parameter & $b$ & $0.656 \pm 0.015$ & \\
Orbital inclination & $i$ & $79.42 \pm 0.39$ & $^\circ$\\
\noalign{\medskip}
Semi-amplitude of the stellar reflex velocity & $K_{\rm 1}$ & $256.9 \pm 2.7$ & \ms\\
Centre-of-mass velocity & $\gamma_{\rm rv1}$ & $20\,787.26 \pm 0.23$ & \ms\\
Offset between HARPS and CORALIE & $\gamma_{\rm HARPS-CORALIE}$ & $19.32 \pm 0.82$ & \ms\\
\noalign{\medskip}
Orbital eccentricity & $e$ & $0.0019^{+ 0.0055}_{- 0.0015}$ & \\
& & $< 0.027$ (3 $\sigma$) & \\
Argument of periastron & $\omega$ & $75^{+ 24}_{- 162}$ & $^\circ$\\
& $e\cos\omega$ & $0.00010^{+ 0.00081}_{- 0.00069}$ & \\
& $e\sin\omega$ & $0.0007^{+ 0.0062}_{- 0.0016}$ & \\
Phase of mid-occultation, having accounted for light travel time & $\phi_{\rm mid-occ.}$ & $0.50006^{+ 0.00052}_{- 0.00044}$ & \\
Occultation duration & $T_{\rm 58}$ & $0.06564^{+0.00043}_{-0.00040}$ & d\\
Duration of occultation ingress $\approx$ duration of occultation egress & $T_{\rm 56} \approx T_{\rm 78}$ \medskip & $0.01358 \pm 0.00054$ & d\\
\noalign{\medskip}
Relative planet-to-star flux at 1.6  \micron & \fhband	& $0.276 \pm 0.044$ & per cent\\
Relative planet-to-star flux at 2.09 \micron & \fnbtwo	& $0.366 \pm 0.067$ & per cent\\
Relative planet-to-star flux at 3.6  \micron & \iracone	& $0.483 \pm 0.025$ & per cent\\
Relative planet-to-star flux at 4.5  \micron & \iractwo	& $0.572 \pm 0.030$ & per cent\\
Relative planet-to-star flux at 5.8  \micron & \iracthree	& $0.65  \pm 0.11$ & per cent\\
Relative planet-to-star flux at 8.0  \micron & \iracfour	& $0.73  \pm 0.12$ & per cent\\
Planet brightness temperature$^{\dagger}$ at 1.6 \micron & $T_{\rm b,1.6}$ & $2750 \pm 130$ & K\\
Planet brightness temperature$^{\dagger}$ at 2.09 \micron & $T_{\rm b,2.09}$ & $2670 \pm 170$ & K\\
Planet brightness temperature$^{\dagger}$ at 3.6 \micron & $T_{\rm b,3.6}$ & $2346 \pm 57$ & K\\
Planet brightness temperature$^{\dagger}$ at 4.5 \micron & $T_{\rm b,4.5}$ & $2273 \pm 64$ & K\\
Planet brightness temperature$^{\dagger}$ at 5.8 \micron & $T_{\rm b,5.8}$ & $2260 \pm 230$ & K\\
Planet brightness temperature$^{\dagger}$ at 8.0 \micron & $T_{\rm b,8.0}$ \medskip & $2260 \pm 250$ & K\\
\noalign{\medskip}
Sky-projected stellar rotation velocity & \vsini & $4.63 \pm 0.27$ & \kms\\
Sky-projected angle between stellar spin and planetary orbit axes & $\lambda$ & $4.1 \pm 5.2$ & $^\circ$\\
\noalign{\medskip}
Star mass & \mstar & $0.969 \pm 0.023$ & \msol\\
Star radius & \rstar & $0.993 \pm 0.018$ & \rsol\\
Star density & \densstar & $0.990 \pm 0.043$ & \denssol\\
Star surface gravity & $\log g_{*}$ & $4.430 \pm 0.012$ & (cgs)\\
Star effective temperature & \teff\ & $5475 \pm 98$ & K\\
Star metallicity & \feh & $0.02 \pm 0.09$ & (dex)\\
\noalign{\medskip}
Planet mass & \mplanet & $1.165 \pm 0.023$ & \mjup\\
Planet radius & \rplanet & $1.383 \pm 0.031$ & \rjup\\
Planet density & \densplanet & $0.440 \pm 0.026$ & \densjup\\
Planet surface gravity & $\log g_{\rm P}$ & $3.133 \pm 0.017$ & (cgs)\\
Semi-major axis & $a$ & $0.01653 \pm 0.00013$ & AU\\
Planet equilibrium temperature$^{\ddagger}$ (full redistribution) & \teqlfull & $2045 \pm 41$ & K\\
Planet equilibrium temperature$^{\ddagger}$ (day side redistribution) & \teqlday & $2432 \pm 49$ & K\\
Planet equilibrium temperature$^{\ddagger}$ (instant re-radiation) & \teqlinst & $2614 \pm 52$ & K\\
\\ 
\hline 
\multicolumn{4}{l}{$^{\dagger}$ We modelled both star and planet as black bodies 
and took account of only the occultation depth uncertainty, which dominates.}\\
\multicolumn{4}{l}{$^{\ddagger}$ 
\teqlf\ $= f^{\frac{1}{4}}T_{\rm eff}\sqrt{\frac{R_*}{2a}}$ where $f$ is the 
redistribution factor, with $f=1$ for full redistribution, $f=2$ for dayside 
redistribution}\\
\multicolumn{4}{l}{and $f=8/3$ for instant re-radiation 
\citep{2011ApJ...729...54C}. We assumed the planet albedo to be zero, $A=0$.}
\end{tabular} 
\end{table*} 

\begin{table}
\caption{Comparing adopted and permuted occultation depths} 
\label{tab:permut}
\begin{tabular}{lll}
\hline
Lightcurve & Adopted depth & Permuted depth\\
\hline
1.6\,\micron\  & $0.00276 \pm 0.00044$  & $0.00270 + 0.00095$\\
2.09\,\micron\ & $0.00366 \pm 0.00067$  & $0.00371 + 0.00035$ \\
3.6\,\micron\  & $0.00483 \pm 0.00025$  & $0.00485 + 0.00018$ \\
4.5\,\micron\  & $0.00572 \pm  0.00030$ & $0.00567 +0.00071 -0.00030$ \\
5.8\,\micron\  & $0.0065 \pm  0.0011$   & $0.00634 \pm 0.00050$ \\
8.0\,\micron\  & $0.0073 \pm  0.0012$   & $0.00824 \pm 0.00077$ \\
\hline
\end{tabular}
\end{table}

\subsection{Stellar activity}

\citet{2010ApJ...708..224H} reported a rotational modulation of the WASP 
lightcurves with a period of $10.5 \pm 0.2$ days and an amplitude of a few mmag. 
This indiciated that WASP-19 is an active star, with the sinusoidal modulation 
being induced by a non-axisymmetric distribution of starspots. 

We determine the \rhk\ activity index of WASP-19 by measuring the weak 
emission in the cores of the Ca {\sc ii} H+K lines 
\citep{1984ApJ...287..769N, 2000A&A...361..265S, 2009A&A...495..959B}. 
The 36 HARPS spectra presented in \citet{2011ApJ...730L..31H} had SNR in the 
range 14--38. We selected the 12 spectra with SNR$>$19 per pixel at 550 nm, as 
the activity level tends to be systematically under- or over-estimated for 
spectra with low SNR. 
By assuming $B-V$ = 0.570, we infer \rhk\ = $-4.50 \pm 0.03$, which is the 
weighted mean and standard deviation of the values determined from individual 
spectra; we used the SNR as the weighting factor. 
This is similar to the value of \rhk\ = $-4.66$ measured by 
\citet{2010ApJ...720.1569K}. It is difficult to judge the level at which the two 
values agree as \citet{2010ApJ...720.1569K} do not provide an uncertainty 
estimate and our uncertainty value is likely to be an underestimate. 

As we know the true stellar rotation period to be $10.5 \pm 0.2$ days from 
rotational modulation, we can use our \rhk\ value to test the 
activity--rotation calibration of \citet{2008ApJ...687.1264M}. 
The calibration suggests a stellar-rotation period of 
$P_{\rm rot} = 12.3 \pm 1.5$ d, which is consistent within errors. 

We considered whether stellar variability could have affected our measured 
occultation depths.
One potential issue is that that the stellar brightness may have varied 
significantly during one or more of the observations. 
However, with observation durations of $\sim$3 hr and a stellar rotation period 
of 10.5 d, the visible portion of the stellar surface will have changed little 
during any one observation. 
To first order, the resulting small impact on the occultation lightcurves can 
be modelled as a linear trend, which will be handled by the trend functions. 
Another concern is that the stellar brightness may have changed significantly 
between the non-simultaneous occultation observations. 
For example, the 3.6-\micron\ data were obtained two months after the 
4.5-\micron\ data and it is the relative measurements at these two wavelengths 
that are the prime diagnostic for the presence of an atmospheric temperature 
inversion. 
Assuming a constant planet brightness, the stellar brightness would need to 
have changed by $\sim$5 per cent to have changed the occultation depth by 
1\,$\sigma$ 
and the amplitude of the modulation of the WASP lightcurves (a few mmag) 
shows that this is very unlikely. 
Thus our derived eclipse depths, and the conclusions on which they depend, are 
insensitive to the variability of WASP-19.

\section{Discussion}

\subsection{Atmosphere model}

We interpret our observations of the hot Jupiter WASP-19b using the exoplanetary 
atmospheric modeling and retrieval method developed in 
\citet{2009ApJ...707...24M, 2010ApJ...725..261M, 2011ApJ...729...41M}. 
We model a plane-parallel atmosphere of WASP-19b observed in thermal emission 
at secondary eclipse. The dayside spectrum of the planet is generated using 
line-by-line radiative transfer, with constraints of hydrostatic equilibrium 
and global energy balance, and includes the dominant sources of infrared 
opacity expected in gaseous atmospheres at high temperature. 
Our sources of opacity include molecular absorption due to H$_2$O, CO, CH$_4$, 
CO$_2$, NH$_3$, TiO, and VO 
\citep{2008ApJS..174..504F, 2005Rothman, 2010Icar..205..674K} 
and H$_2$--H$_2$ collision-induced absorption 
\citep{1997A&A...324..185B, 2002A&A...390..779B}. 
The concentrations of the species and the pressure--temperature ($P$--$T$) 
profile constitute the free parameters in the model \citep{2009ApJ...707...24M}. 
We explore the parameter space of the model using a Markov-chain Monte Carlo 
scheme 
\citep[see][]{2010ApJ...725..261M, 2011ApJ...729...41M}, and 
constrain regions of parameter space consistent with the measured planet-to-star 
flux density ratios at different levels of fit. 
Our goal is to constrain the existence of a possible temperature 
inversion, the dayside-to-nightside redistribution efficiency, the 
concentrations of the different molecular species, and the C/O 
ratio \citep[e.g.][]{2011Natur.469...64M} in the dayside atmosphere of 
WASP-19b. In what follows, we discuss model solutions that explain the data 
within the 1-$\sigma$ observational uncertainties, as shown in 
Fig.~\ref{fig:atmos}. 

\subsection{Temperature inversion}

The data indicate the lack of a strong temperature  inversion in the dayside 
atmospheres of WASP-19b. The observations and two models fitting the data are 
shown in Fig.~\ref{fig:atmos}. 
The lack of a temperature  inversion in WASP-19b is evident from the data even 
without detailed modeling. 
Firstly, the brightness temperature in the 4.5 $\micron$ channel, 
$T_{\rm b,4.5}$, is lower than that in the 3.6 $\micron$ channel, 
$T_{\rm b,3.6}$ (Table~\ref{tab:sys-params}). 
The presence of a temperature  inversion is often indicated by a 
$T_{\rm b,4.5}$ value considerably higher than the $T_{\rm b,3.6}$ value, due 
to strong CO emission and some H$_2$O emission in the 4.5 $\micron$ channel 
\citep{2008ApJ...678.1436B, 2008ApJ...678.1419F, 2010ApJ...725..261M}.
Secondly, the brightness temperatures at 1.6 and 2.09 \micron\ are larger than 
those in all four IRAC channels, which are at longer wavelengths. Since 
the bands at 1.6 and 2.09 \micron\ are windows in molecular opacity, they are 
expected to probe temperatures in deeper layers of the atmosphere compared to 
any of the IRAC channels. 
Therefore, the high temperatures in the 1.6 and 2.09 \micron\ bands compared to 
all the IRAC channels imply temperature decreasing outwards in the atmosphere, 
and hence the absence of a temperature inversion or, at most, the presence of 
one too weak to be detectable with the available data. 
Two model $P$--$T$ profiles without 
temperature inversions and the corresponding model spectra are shown in 
Fig.~\ref{fig:atmos}. All the IRAC data can be explained by molecular 
absorption in the atmosphere due to the temperature decreasing outwards.  

The lack of a temperature  inversion in WASP-19b offers a new constraint on 
existing classification schemes of irradiated giant exoplanets. WASP-19b falls 
in the category of highly irradiated hot Jupiters which have been predicted to 
host temperature  inversions due to TiO and VO, assuming solar abundances 
\citep{2008ApJ...678.1419F} - the so called `TiO/VO hypothesis'. However, our 
finding of a lack of a strong temperature inversion in WASP-19b implies that 
either TiO and VO are depleted or an entirely different process is at play. If 
the composition is oxygen-rich, the lack of a temperature inversion in WASP-19b 
can be explained if TiO and VO are depleted in the upper atmosphere due to 
gravitational settling, which can be significant if the vertical mixing is weak 
\citep{2009ApJ...699.1487S}. On the other hand, if the composition is 
carbon-rich, 
TiO and VO would be naturally scarce \citep{2011ApJ...743..191M}. 

\citet{2010ApJ...720.1569K} posit that the high UV flux impinging on those 
planets orbiting chromospherically active stars could destroy the high-opacity, 
high-altitude compounds that would otherwise lead to temperature inversions.  
With \rhk\ = $-4.50 \pm 0.03$, WASP-19 has a similar activity level to that of 
the handful of other stars around which hot Jupiters without temperature 
inversions (\rhk\ = $-4.9$ to $-4.5$) are known to orbit. 
Those planets thought to have inversions orbit quieter stars, with 
\rhk\ = $-5.4$ to $-4.9$. 
Thus, our finding that WASP-19b has no inversion supports the hypothesis of 
\citet{2010ApJ...720.1569K} and usefully adds to the handful of such systems 
known. 
{\it Spitzer} routinely measures the thermal emission of planets at 3.6 and 
4.5 \micron. 
For planets with temperature inversions, CO and water switch from absorption to 
emission, resulting in a higher flux in the 4.5 \micron\ band, in which these 
molecules have features. 
\citet{2010ApJ...720.1569K} proposed a model-independent, empirical metric for 
classifying hot Jupiters, which we denote with $\zeta$.
This is the gradient of the measurements at 3.6 and 4.5 \micron, i.e. 
$(\iractwo - \iractwo) / 0.9 \micron$, minus the gradient of the 
black body that is the best-fit to the two measurements. 
A positive $\zeta$-value would suggest an inverted atmosphere and a strongly 
negative $\zeta$-value would indicate a non-inverted atmosphere; 
\citet{2010ApJ...720.1569K} 
suggest a delineation around $\zeta = -0.05$ per cent \micron$^{-1}$.  
For WASP-19b we calculated $\zeta = -0.031 \pm 0.043$ per cent 
\micron$^{-1}$. 
This value is at the border between inversion and no-inversion and therefore is 
consistent with our finding that WASP-19b does not have a strong inversion. 

We note that in their activity-inversion plot (their figure~5), 
\citet{2010ApJ...720.1569K} omit XO-3 but include TrES-4, 
HAT-P-7 and WASP-18, all of which have $\teff \gtrsim 6200$. 
XO-3b has a temperature inversion and XO-3 has a \rhk\ index indicative of 
activity, which seems to contradict the activity-inversion hypothesis.
The other three planets have inversions and orbit quiet stars. 
\citet{2010ApJ...720.1569K} concluded that in fact XO-3 is likely to be 
chromospherically quiet, based on a visual inspection of their spectrum  
and having noted that the \rhk\ calibration is unreliable for stars with $\teff 
\gtrsim 6200$.
Perhaps then TrES-4, HAT-P-7 and WASP-18 are also suspect since they also 
have $\teff \gtrsim 6200$. 

\subsection{Atmospheric composition}
\begin{figure}
\includegraphics[width=0.5\textwidth]{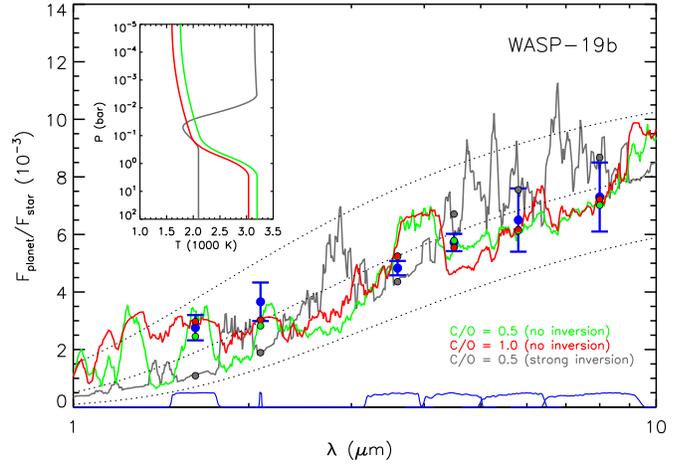}
\caption{Spectral energy distribution of WASP-19b relative to that of its host 
star. The blue dots are our fitted planet-to-star flux density ratios. 
The observation-band transmission curves are shown along the abscissa. 
The planet-to-star flux density ratios indicate the absence of a strong 
temperature inversion. 
Two model-atmosphere spectra are shown, each of which fits the data and lacks a 
temperature inversion. 
One model has a C/O ratio of 0.5 and is shown as a green line. 
The other has a C/O ratio of 1.0 is depicted by a red line. 
The corresponding band-integrated model fluxes are shown as dots with the same 
colour scheme. 
The models' pressure--temperature profiles down to 100 bar are shown in the 
inset. 
Assuming zero albedo, the maximum dayside-to-nightside redistribution 
efficiency is 20 per cent for the C-rich model and 34 per cent for the 
O-rich model. 
The three dotted lines show planetary black bodies with temperatures of 1800, 
2250 and 2900 K (chosen to give a sense of scale to the spectra) divided by 
a stellar spectrum. 
\label{fig:atmos}}
\end{figure}

We find that the observations can be explained by models with oxygen-rich 
as well as carbon-rich compositions. 
The absorption in the near-IR (1.6 and 2.09 \micron) bands is minimal due to 
the lack of major molecular features.
The constraints on the composition come primarily from the IRAC data, which 
together encompass features of CO, H$_2$O, CH$_4$, and CO$_2$. 
The near-IR data, however, are critical to constraining the temperature of the 
lower atmosphere and thus are key in anchoring the model-atmosphere spectra to 
the measured SED. 
Two models with different C/O ratios, C/O = 0.5 (oxygen-rich) and C/O = 1 
(carbon-rich), are shown in Fig.~\ref{fig:atmos}. 
As demonstrated in \citet{2011ApJ...743..191M}, CO is a dominant carbon-bearing 
molecule in both C-rich and O-rich regimes. Consequently, the 4.5 $\micron$ 
absorption in both models in Fig.~\ref{fig:atmos} is caused primarily by CO 
absorption; in the O-rich model CO$_2$ contributes additional absorption in 
this channel. 
The absorption in the 3.6, 5.8, and 8.0 micron IRAC channels in the O-rich model 
is caused primarily by H$_2$O absorption, whereas absorption in the C-rich model 
is caused by a combination of H$_2$O and CH$_4$ absorption; H$_2$O is depleted 
by a factor of 100 and CH$_4$ is enhanced by a factor of 1000 with respect to 
the O-rich model, both of which are chemically feasible 
\citep{2011ApJ...743..191M}. 
The principle difficulty in differentiating between the two models with the 
current data are the large uncertainties in the 5.8 and 8.0 $\micron$ IRAC data. 
For example, a high 5.8 $\micron$ point would indicate low water absorption, 
and hence high C/O, as demonstrated in \citet{2011Natur.469...64M}. 
New observations in the near infrared can 
differentiate between spectra from the carbon-rich and oxygen-rich compositions 
\citep[e.g.][]{2011Natur.469...64M}. 
The water abundance can be measured via transmission spectoscopy of the 1.4 
\micron\ water band using the G141 grism of HST/WFC3; these observations were 
recently peformed for WASP-19b by \citet{2009hst..prop12181D}. 
We could measure, or at least place useful constraints on, the TiO abundance 
with ground-based occultation observations in the $z$ and $J$ bands.

\subsection{Orbital eccentricity}

\begin{figure}
\includegraphics[width=0.5\textwidth]{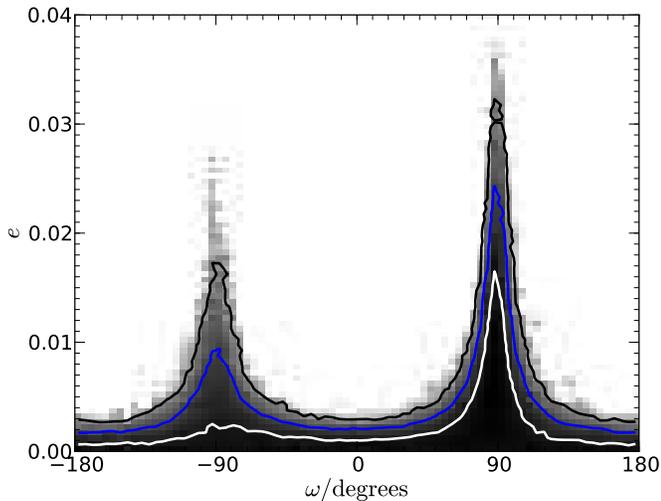}
\caption{
The MCMC posterior distributions of $e$ and $\omega$.
The white, blue and black contours are, respectively, the 1-, 2- and 3-$\sigma$ 
confidence limits. 
The shading of each bin is proportional to the logarithm of the number of MCMC 
steps within. 
The available data exclude large values of $e$ for any orbital orientation, 
and only very small values of $e$ are permitted unless $|\omega| \approx 90$. 
\label{fig:eom}}
\end{figure}

For a circular orbit, mid-occultation occurs half an orbital period after 
mid-transit. 
We find the occultation to occur only $4^{+35}_{-30}$ s later than this and 
constrain both \ecos\ and \esin\ to a small region around zero. 
Hence, the orbit is very nearly circular, though the available 
data do permit a small, non-zero eccentricity providing that the major axis of
the orbit is near-aligned with our line of sight, such that the occulation time 
is not affected (i.e. $|\omega| \approx 90$; Figure~\ref{fig:eom}). 
We place a 3-$\sigma$ upper limit on eccentricity of $e < 0.027$. 

Hot Jupiters are considered to have been moved
inwards to close orbits through planet--planet scattering or
by the Kozai mechanism and tidal circularisation 
\citep[e.g.:][]{2011Natur.473..187N,2011ApJ...735..109W,2011A&A...533A...7B}. 
However, for the very shortest-period
systems, such as WASP-19b, it is unlikely that they could have
been moved directly to their current orbit, since that would
have required careful fine-tuning to avoid destruction
by collision with the star.  Thus most likely WASP-19b was
first moved to an orbit near $\sim$\,2 Roche radii and has
since spiralled inwards through tidal orbital decay 
(see \citealp{2011ApJ...732...74G} and the discussion of WASP-19b specifically 
in \citealp{2011ApJ...730L..31H}). 

From tidal theory, the circularisation of a hot Jupiter's orbit is thought 
to proceed much faster than the infall, and this is consistent with the 
the observation that hot Jupiters tend to be in circular orbits. 
Thus the suggestion
that WASP-19b has undergone significant tidal decay, from
$\sim$\,2 Roche radii to the current 1.2 Roche radii, leads
to the expectation that the current eccentricity will be 
essentially zero, in line with our results.

\section*{Acknowledgments}
This work is based on observations made with the Spitzer 
Space Telescope, which is operated by the Jet Propulsion Laboratory, 
California Institute of Technology under a contract with NASA. 
We thank N. P. Gibson for providing the HAWK-I 2.09-\micron\ lightcurve. 


\label{lastpage}

\end{document}